\documentclass[onecolumn,prd,floatfix,preprintnumbers,nofootinbib,%
superscriptaddress]{revtex4}
\usepackage[T1]{fontenc}
\usepackage{lmodern} 
\usepackage[utf8]{inputenc} 
\usepackage{mathrsfs}
\usepackage{amssymb}	
\usepackage{amsmath}
\usepackage{graphicx}
\usepackage{xcolor}
\usepackage{placeins}
\usepackage[normalem]{ulem}
\usepackage{stmaryrd}

\definecolor{lcolor}{rgb}{0.5,0,0}
\definecolor{citcolor}{rgb}{0,0.3,0.0}
\definecolor{coloryksi}{rgb}{0.5,0.0,0.0}
\definecolor{colorkaksi}{rgb}{0.0,0.0,0.5}
\definecolor{colorkolme}{rgb}{0.0,0.3,0.1}

\usepackage{mathtools}
\usepackage{physics}
\usepackage{tensor}
\usepackage{cancel}
\usepackage{mathrsfs}

\newcommand{\beq}{\begin{equation}}
\newcommand{\eeq}{\end{equation}}

\usepackage[breaklinks,colorlinks,urlcolor=blue,citecolor=citcolor,linkcolor=lcolor]{hyperref}

\usepackage{tikz}
\usetikzlibrary{snakes}

\newcounter{diag}
\newcounter{subdiag}[diag]
\begin{document}

\author{Yair Mulian}
\affiliation{Institute of Physics, Academia Sinica, Taipei, 11529, Taiwan}

\title{The Magnus expansion for non-Hermitian Hamiltonians}


\begin{abstract}
The Magnus expansion offers a method to express a time-ordered exponential as an ordinary operatorial exponential. This representation has advantageous theoretical properties, while still solving the original differential equation. For any finite dimensional Hermitian Hamiltonians, the standard Magnus expansion guarantees a manifestly unitary representation. However, this property is no longer preserved if the Hamiltonian is infinite dimensional or non-Hermitian. In this work, we derive a generalized expansion that maintains the property of unitarity for all bounded finite dimensional Hamiltonians.
\end{abstract}

\maketitle

\tableofcontents

\section{Introduction}
According to the Schrödinger equation \cite{schro}, the time evolution operator $\hat{U}(t,t_{0})$ that governs the time evolution of an initially prepared wave-function $\left|\psi(t_{0})\right\rangle $ satisfies
\begin{equation}\label{schro}
\frac{d\hat{U}(t,t_{0})}{dt}\,=\,-i\hat{H}(t)\hat{U}(t,t_{0}).
\end{equation}
In case that the Hamiltonian $\hat{H}(t)$ contains a non-trivial time dependence, the familiar solution for (\ref{schro}) is known as the Dyson series \cite{freema1,freema2,landau,Peskin},
\begin{equation}\label{dys}
\hat{U}(t,\,t_{0})\,=\,\hat{T}\exp\left[-i\int_{t_{0}}^{t}dt^{\prime}\,\hat{H}(t^{\prime})\right].
\end{equation}
This way of expressing the solution involves the time ordering operator $\hat{T}$,
\begin{equation}
\hat{T}\left[\hat{H}(t^{\prime})\,\hat{H}(t)\right]\,\equiv\,\Theta(t^{\prime}-t)\hat{H}(t^{\prime})\,\hat{H}(t)\,+\,\Theta(t-t^{\prime})\hat{H}(t)\,\hat{H}(t^{\prime}),
\end{equation}
where $\Theta(t^{\prime}-t)$ is the Heaviside theta function. The result a series in which each term involves higher order of integrated iterations,
\begin{equation}\label{uexp1}
\hat{U}(t,\,t_{0})\,=\,\hat{\mathbf{1}}\,-\,i\int_{t_{0}}^{t}dt^{\prime}\:\hat{H}(t^{\prime})\,-\,\int_{t_{0}}^{t}dt^{\prime}\int_{t_{0}}^{t^{\prime}}dt^{\prime\prime}\,\hat{H}(t^{\prime})\,\hat{H}(t^{\prime\prime})\,+\,\ldots
\end{equation}
The drawback of such a representation is that the algebraic rules for handling a time-dependent exponential is not as convenient as the corresponding regular exponential.  For example, according the Baker–Campbell–Hausdorff formula \cite{Gilmore}, for possibly noncommutative bounded matrices $\hat{A}$ and $\hat{B}$ on Banach space, 
\begin{equation}
\exp\hat{A}\,\exp\hat{B}\,=\,\exp\hat{C},
\end{equation}
with 
\begin{equation}\label{bkhseries}
\hat{C}\,=\,\hat{A}+\hat{B}+\frac{1}{2}\left[\hat{A},\hat{B}\right]+\frac{1}{12}\left[\hat{A},\left[\hat{A},\hat{B}\right]\right]-\frac{1}{12}\left[\hat{B},\left[\hat{A},\hat{B}\right]\right]+\ldots 
\end{equation}
However, no such a simple construction rule is known if the exponential is replaced by a time ordered one. Moreover, the time-ordering obscures the property of unitarity, as it might not be possible to act with the conjugation operator straightforwardly.  An insightful idea that has been put forward by Wilhelm Magnus (1907–1990) is to construct an expansion that is equivalent to (\ref{dys}) as a standard exponential. In terms of mathematical justification, the existence of an equivalent manifestly unitary representation is guaranteed by the Stone theorem \cite{stone} only for self-adjoint\footnote{Since in this work the underlying space is assumed to be of finite dimension, self-adjoint and Hermitian are equivalent concepts.} Hamiltonians, $\hat{H}^{\dagger}(t)=\hat{H}(t)$.
In his seminal work \cite{Magnus}, Magnus was able to show that under certain convergence assumptions there exist an equivalent representation for Hermitian Hamiltonians (\ref{dys}),
\begin{equation}\label{mgnu}
\hat{U}(t,\,t_{0})\,=\,\exp\left[-i\int_{t_{0}}^{t}dt^{\prime}\,\hat{\Omega}(t^{\prime})\right].
\end{equation}
The generator $\hat{\Omega}(t)$ is the Magnus series generator (the dependence on the parameter $t_{0}$ is kept implicit) which denotes an infinite series,
\begin{equation}
\hat{\Omega}(t)\,=\,\sum_{n=1}^{\infty}\hat{\Omega}_{n}(t).
\end{equation}
As shown in \cite{Blanes}, the first few terms in this perturbative series read
\begin{equation}\begin{split}\label{gener}
&\hat{\Omega}(t)\,=\,\hat{H}(t)\,+\,\frac{1}{2}\int_{t_{0}}^{t}dt^{\prime}\:\left[\hat{H}(t),\,\hat{H}(t^{\prime})\right]\\
&+\,\frac{1}{6}\int_{t_{0}}^{t}dt^{\prime}\int_{t_{0}}^{t^{\prime}}dt^{\prime\prime}\:\left(\left[\hat{H}(t),\,\left[\hat{H}(t^{\prime}),\,\hat{H}(t^{\prime\prime})\right]\right]+\left[\hat{H}(t^{\prime\prime}),\,\left[\hat{H}(t^{\prime}),\,\hat{H}(t)\right]\right]\right)+\ldots 
\end{split}\end{equation}
One of the main advantages of the Magnus expansion is that the truncated series very often shares important qualitative properties with the exact solution. For instance, in classical mechanics the symplectic character of the time evolution is preserved at every order of approximation. A more detailed analysis of the capacity of expansion (\ref{mgnu}) to provide an equivalent representation to (\ref{dys}) is discussed in \cite{Blanes,pedag,fard,iserl}. The convergence properties of the Magnus expansion is discussed in \cite{casas}.

\section{Unitarizing the Dyson expansion}
At this point, it is clear that establishing an equivalence between the Dyson and Magnus cannot be done for Hamiltonian that are not Hermitian. Our intention here is to extend the capacity of such an equivalence to hold for any type of Hamiltonian, for which possibly
 \begin{equation}
\hat{H}^{\dagger}(t)\,\neq\,\hat{H}(t).
\end{equation}
The first step in order to achieve that requires to adopt the approach that has been developed in \cite{mulian}. This amounts in proceeding by replacing (\ref{dys}) with an alternative construction that is manifestly unitary. Our approach is achieving that by unitarization of the time evolution operator, that is carried by the introduction of a dynamical normalization operator,
\begin{equation}\label{unitrz}
\hat{U}\,\longrightarrow\,\mathcal{\hat{P}}=\sqrt{\hat{U}^{\dagger-1}\,\hat{U}^{-1}}\,\hat{U}.
\end{equation}
This decomposition is analogous to the polar decomposition, which always exists and its positive root is unique. Based on the unitarization procedure, it has been shown in \cite{mulian} that construction (\ref{unitrz}) provides us with a revised time evolution operator that reduces to the standard one for Hermitian Hamiltonians. More explicitly, with the new evolution operator, the time evolved state is computed via the action of
\begin{equation}\label{pitaron}
\mathcal{\hat{P}}(t,\,t_{0})\,\equiv\,\hat{\mathcal{N}}(t,\,t_{0})\,\hat{U}(t,\,t_{0}),\qquad\qquad\left|\Psi(t)\right\rangle \,=\,\mathcal{\hat{P}}(t,\,t_{0})\left|\Psi(t_{0})\right\rangle.
\end{equation}
The new operator is $\mathcal{N}$ introduced as the positive root of its square which is by itself a positive operator,
\begin{equation}\label{definorm}
\mathcal{\hat{N}}(t,\,t_{0})\,\equiv\,\sqrt{\hat{U}^{\dagger-1}(t,\,t_{0})\,\hat{U}^{-1}(t,\,t_{0})}\,.
\end{equation}
By construction, the solution $\mathcal{\hat{P}}(t,\,t_{0})$ manifestly preserves the exactness of unitarity at all orders and at all times,
\begin{equation}\label{fullunitar}
\mathcal{\hat{P}}^{\dagger}(t,\,t_{0})\,\mathcal{\hat{P}}(t,\,t_{0})=\hat{U}^{\dagger}(t,\,t_{0})\,\hat{U}^{\dagger-1}(t,\,t_{0})\,\hat{U}^{-1}(t,\,t_{0})\,\hat{U}(t,\,t_{0})\,=\,\boldsymbol{\hat{1}}.
\end{equation}
Thus, for the new construction it is manifestly apparent that $\mathcal{\hat{P}}^{\dagger}(t,\,t_{0})\,=\,\mathcal{\hat{P}}^{-1}(t,\,t_{0})\,$.
The expression for the inverse of $\hat{U}$ can be uniquely determined by the condition $\hat{U}^{-1}(t,\,t_{0})\,\hat{U}(t,\,t_{0})\,=\,\hat{\boldsymbol{1}}$,
\begin{equation}\label{uinv}
\hat{U}^{-1}(t,t_{0})\,=\,\hat{\mathbf{1}}+i\int_{t_{0}}^{t}dt^{\prime}\:\hat{H}(t^{\prime})-\left(\int_{t_{0}}^{t}dt^{\prime}\,\hat{H}(t^{\prime})\right)^{2}+\int_{t_{0}}^{t}dt^{\prime}\int_{t_{0}}^{t^{\prime}}dt^{\prime\prime}\:\hat{H}(t^{\prime})\,\hat{H}(t^{\prime\prime})\,+\,\ldots
\end{equation}
After insertion back to (\ref{definorm}) along with the expansion of the square root a perturbative version can be obtained, which has been computed in \cite{mulian}.

\section{Generalizing the Magnus expansion}
In an analogous way to complex numbers, any bounded non-Hermitian Hamiltonian \cite{nonhe1,nonhe2,nonhe3,nonhe4} can be expressed as a sum of an Hermitian and anti-Hermitian parts,
\begin{equation}\label{rep}
\hat{H}(t)=\hat{\mathcal{H}}(t)+i\hat{\mathcal{J}}(t),\qquad\qquad\hat{H}^{\dagger}(t)=\hat{\mathcal{H}}(t)-i\hat{\mathcal{J}}(t).
\end{equation}
with $\hat{\mathcal{H}}(t)=\hat{\mathcal{H}}^{\dagger}(t)$ and $\hat{\mathcal{J}}(t)=\hat{\mathcal{J}}^{\dagger}(t)$. Since the Hamiltonian is assumed to be expressed by a diagonalizable matrix, its two components must commute for any given value of $t$, 
\begin{equation}\label{inv}
\left[\hat{\mathcal{H}}(t),\hat{\mathcal{J}}(t)\right]\,=\,0.
\end{equation}
The representation (\ref{rep}) can also be inverted,
\begin{equation}\label{hermpart}
\hat{\mathcal{H}}(t)=\frac{1}{2}\left(\hat{H}(t)+\hat{H}^{\dagger}(t)\right),\quad\quad\hat{\mathcal{J}}(t)=\frac{1}{2i}\left(\hat{H}(t)-\hat{H}^{\dagger}(t)\right).
\end{equation}
Our intention in this part is to compute an analogous expansion for (\ref{mgnu}) based on the operator (\ref{pitaron}), which no longer involve the time ordering operation. As expected, the new expansion must to reduce back to the original expansion if the Hamiltonian is bounded self-adjoint, as in that case $\mathcal{\hat{N}}=\hat{\boldsymbol{1}}$. The unitarized time evolution operator is related to the generator $\hat{\Sigma}(t)$ via
\begin{equation}\label{nhmag}
\hat{\mathcal{P}}(t,\,t_{0})\,=\,\exp\left[-i\int_{t_{0}}^{t}dt^{\prime}\,\hat{\Sigma}(t^{\prime})\right],\quad\qquad\hat{\Sigma}(t)\,=\,\sum_{n=0}^{\infty}\hat{\Sigma}_{n}(t).
\end{equation}
In accordance with the property of unitarity $\hat{\mathcal{P}}^{\dagger}(t,\,t_{0})=\hat{\mathcal{P}}^{-1}(t,\,t_{0})$, and therefore, $\hat{\Sigma}^{\dagger}(t)=\hat{\Sigma}(t)$. The differential equation analogous to (\ref{schro}) no longer involves the imaginary component,
\begin{equation}\label{newschro}
\frac{d\hat{\mathcal{P}}(t,t_{0})}{dt}\,=\,-i\hat{\mathcal{H}}(t)\hat{\mathcal{P}}(t,t_{0}).
\end{equation}
Since the Hamiltonian is assumed to be given by a bounded operator, the conjugation and inverse operations involved in definition (\ref{definorm}) can be iterated and $\hat{\mathcal{N}}$ can be alternatively represented as
\begin{equation}\label{revdefn}
\hat{\mathcal{N}}(t,\,t_{0})=\left(\sqrt{\hat{U}(t,\,t_{0})\,\hat{U}^{\dagger}(t,\,t_{0})}\right)^{-1}.
\end{equation}
After applying the conjugation on eq. (\ref{mgnu}),
\begin{equation}
\hat{U}^{\dagger}(t,t_{0})\,=\,\left(\exp\left[-i\int_{t_{0}}^{t}dt^{\prime}\,\hat{\Omega}(t^{\prime})\right]\right)^{\dagger}\,=\,\exp\left[i\int_{t_{0}}^{t}dt^{\prime}\,\hat{\Omega}^{\dagger}(t^{\prime})\right].
\end{equation}
Based on (\ref{revdefn}), the product of the two involved exponentials can be rewritten as
\begin{equation}
\exp\left[-i\int_{t_{0}}^{t}dt^{\prime}\,\hat{\Omega}(t^{\prime})\right]\,\exp\left[i\int_{t_{0}}^{t}dt^{\prime}\,\hat{\Omega}^{\dagger}(t^{\prime})\right]\,=\,\exp\left[\int_{t_{0}}^{t}dt^{\prime}\,\hat{\Xi}(t^{\prime})\right],
\end{equation}
Since the operator $\hat{\mathcal{N}}$ is by construction Hermitian, that implies that the generator $\hat{\Xi}$ must also be an Hermitian, $\hat{\Xi}^{\dagger}(t)=\,\hat{\Xi}(t)$. With the aid (\ref{bkhseries}) one arrives at the expression
\begin{equation}\begin{split}\label{xiformula}
&\hat{\Xi}(t)\,=\,-i\left(\hat{\Omega}(t)-\hat{\Omega}^{\dagger}(t)\right)+\frac{1}{2}\int_{t_{0}}^{t}dt^{\prime}\left[\hat{\Omega}(t),\hat{\Omega}^{\dagger}(t^{\prime})\right]\\
&-\frac{i}{12}\int_{t_{0}}^{t}dt^{\prime}dt^{\prime\prime}\left(\left[\hat{\Omega}(t),\left[\hat{\Omega}(t^{\prime}),\hat{\Omega}^{\dagger}(t^{\prime\prime})\right]\right]+\left[\hat{\Omega}^{\dagger}(t),\left[\hat{\Omega}(t^{\prime}),\hat{\Omega}^{\dagger}(t^{\prime\prime})\right]\right]\right)+\ldots 
\end{split}\end{equation}
Following definition (\ref{revdefn}) we arrive at the positive root,
\begin{equation}\label{resultN}
\hat{\mathcal{N}}(t,\,t_{0})\,=\,\exp^{-1}\left[\frac{1}{2}\int_{t_{0}}^{t}dt^{\prime}\,\hat{\Xi}(t^{\prime})\right]\,=\,\exp\left[-\frac{1}{2}\int_{t_{0}}^{t}dt^{\prime}\,\hat{\Xi}(t^{\prime})\right].
\end{equation}
Based on definition (\ref{pitaron}), after introducing (\ref{mgnu}) and (\ref{resultN}):
\begin{equation}
\hat{\mathcal{P}}(t,\,t_{0})\,=\,\exp\left[-\frac{1}{2}\int_{t_{0}}^{t}dt^{\prime}\,\hat{\Xi}(t^{\prime})\right]\exp\left[-i\int_{t_{0}}^{t}dt^{\prime}\,\hat{\Omega}(t^{\prime})\right].
\end{equation}
Finally, in terms of the Magnus representation (\ref{nhmag}), the associated generator is expressed with the aid of (\ref{bkhseries}) as
\begin{equation}\begin{split}\label{expan}
&-i\hat{\Sigma}(t)\,=\,-i\hat{\Omega}(t)-\frac{1}{2}\hat{\Xi}(t)+\frac{i}{4}\int_{t_{0}}^{t}dt^{\prime}\left[\hat{\Xi}(t),\hat{\Omega}(t^{\prime})\right]\\
&+\frac{i}{48}\int_{t_{0}}^{t}dt^{\prime}dt^{\prime\prime}\left(2\left[\hat{\Omega}(t),\left[\hat{\Omega}(t^{\prime}),\hat{\Xi}(t^{\prime\prime})\right]\right]+\left[\hat{\Xi}(t),\left[\hat{\Omega}(t^{\prime}),\hat{\Xi}(t^{\prime\prime})\right]\right]\right)+\ldots 
\end{split}\end{equation}

\section{Example}
\begin{figure}
\center
  \includegraphics[scale=0.6]{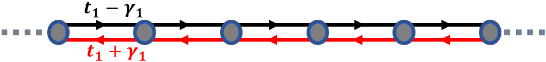}
  \caption{The non-Hermitian skin effect (NHSE) Hamiltonian. At each site the particle can either transit to its left or right nearby site with an unequal time dependent probabilities.\label{skin}}
\end{figure}

In this part we demonstrate with a simple model how our proposed unitarized Magnus expansion is practically applied for an actual time dependent non-Hermitian quantum system. As out toy model, one can use a generalized Hatano-Nelson model (also known as the non-Hermitian skin effect) \cite{nhse1,nhse2} in which the transition amplitudes are no longer time independent. In this model a macroscopic number of particles are initially prepared localized at a boundary of a lattice of size $l$, $\left|\psi_{0}\right\rangle $. These particles are allowed to propagate on the lattice with unequal probabilities (see Figure \ref{skin}) according to the Hamiltonian
\begin{equation}
\hat{H}(t)\,=\,\sum_{i=1}^{l-1}\left[\left(\tau_{i}-\gamma_{i}(t)\right)a_{i}^{\dagger}a_{i+1}+\left(\tau_{i}+\gamma_{i}(t)\right)a_{i+1}^{\dagger}a_{i}\right].
\end{equation}
In the above construction $a_{i}$ and $a_{i}^{\dagger}$ are the creation and annihilation operators of a particle in the $i$th unit cell, $\tau_{i}\in\mathbb{R}$ is independent of time, while $\gamma_{i}\in\mathbb{R}$ is time dependent. As one can observe, due to the finite size of the lattice and the asymmetric nature of the probabilities for a left/right transition ($t_{i}+\gamma_{i}$ vs $t_{i}-\gamma_{i}$), the Hamiltonian is non-Hermitian $\hat{H}^{\dagger}\,\neq\,\hat{H}$. In fact,
\begin{equation}
\hat{H}(t)-\hat{H}^{\dagger}(t)\,=\,2\sum_{i=1}^{l-1}\gamma_{i}(t)\left(a_{i+1}^{\dagger}a_{i}-a_{i}^{\dagger}a_{i+1}\right).
\end{equation}
The standard approach based on (\ref{dys}), will regard the time evolution of this system as a dissipative system, such that $\Delta t\rightarrow\infty$ implies $\left|\psi(t)\right\rangle \rightarrow0$. Since the evolution is not unitary the norm of the WF is not preserved over time and the implementation of the regular Magnus expansion is inadequate in this case. On the other hand, in our proposed formalism the probability after the each step is getting re-distributed in such a way that unitarity is always preserved manifestly, allowing the Born's probabilistic interpretation. Based on substitution of (\ref{xiformula}) inside (\ref{expan}), one finds that
\begin{equation}\begin{split}\label{avera}
\hat{\Sigma}(t)\,=\,\frac{1}{2}\left(\hat{\Omega}(t)+\hat{\Omega}^{\dagger}(t)\right)\,\approx\,\hat{\mathcal{H}}(t).
\end{split}\end{equation}
In the above result we kept for simplicity only the first non-trivial term in the expansion. In accordance with definition (\ref{hermpart}), the Hermitian part of the Hamiltonian is given by
\begin{equation}
\hat{\mathcal{H}}\,=\,\sum_{i=1}^{l-1}\tau_{i}\left(a_{i}^{\dagger}a_{i+1}+a_{i+1}^{\dagger}a_{i}\right).
\end{equation}
Thus, to the first order in the perturbative expansion of the generator, the time evolved wave function $\left|\psi(t)\right\rangle $ is related to the initially prepared WF $\left|\psi_{0}\right\rangle $ via
\begin{equation}
\left|\psi(t)\right\rangle \,=\,\hat{\mathcal{P}}(t,\,t_{0})\left|\psi_{0}\right\rangle \,=\,\exp\left[-i\Delta t\,\hat{\mathcal{H}}\right]\left|\psi_{0}\right\rangle .
\end{equation}

\section{Conclusions}
Our approach has been shown to enlarge the applicability of the Magnus expansion to a broader range of quantum systems. By using expansion (\ref{nhmag}) the feature of unitary becomes universal, even for non-Hermitian Hamiltonians. As a concluding remark, we would like to mention that the new expansion is fully compatible with the familiar expansion for the ordinary Hermitian Hamiltonians. This can be noted by taking the limit $\hat{\mathcal{J}}(t)\rightarrow0$, that implies $\hat{\Xi}(t)=0$, and according to (\ref{avera})
\begin{equation}
\exp\left[-i\int_{t_{0}}^{t}dt^{\prime}\,\hat{\Sigma}(t^{\prime})\right]\rightarrow\exp\left[-i\int_{t_{0}}^{t}dt^{\prime}\,\hat{\Omega}(t^{\prime})\right].
\end{equation}

\section*{Acknowledgements}
Y.M. would like to thank Prof. A. Ramallo for introducing to him the subject of Magnus expansion at the first time. Y.M. thanks Prof. F. Casas for his comments on the manuscript and for the opportunity to participate in the workshop \textit{"70 years of Magnus series"} in which the idea of this work has been initiated.

\section*{Declarations}
\textit{\textbf{Conflict of interest:}} The author have no relevant financial or non-financial interests to disclose. No funding was received for conducting this study.\\\\
\textit{\textbf{Data availability:}} Data sharing not applicable to this article as no datasets were generated or analysed during the current study.\\\\
\textit{\textbf{Open Access:}} This article is licensed under a Creative Commons Attribution 4.0 International License, which permits use, sharing, adaptation, distribution and reproduction in any medium or format, as long as you give appropriate credit to the original author(s) and the source, provide a link to the Creative Commons licence, and indicate if changes were made. The images or other third party material in this article are included in the article’s Creative Commons licence, unless indicated otherwise in a credit line to the material. If material is not included in the article’s Creative Commons licence and your intended use is not permitted by statutory regulation or exceeds the permitted use, you will need to obtain permission directly from the copyright holder. To view a copy of this licence, visit \url{http://creativecommons.org/licenses/by/4.0/}


\begin{thebibliography}{9}

\bibitem{schro}
Schrödinger, E. (1926). An undulatory theory of the mechanics of atoms and molecules. Physical review, 28(6), 1049.

\bibitem{freema1}
Dyson, F. J. (1949). The radiation theories of Tomonaga, Schwinger, and Feynman. Physical Review, 75(3), 486.

\bibitem{freema2}
Dyson, F. J. (1949). The S matrix in quantum electrodynamics. Physical Review, 75(11), 1736.

\bibitem{landau}
Landau, L. D., \& Lifshitz, E. M. (2013). Quantum mechanics: non-relativistic theory (Vol. 3). Elsevier.

\bibitem{Peskin}
Schwartz, M. D. (2014). Quantum field theory and the standard model. Cambridge University press.

\bibitem{Gilmore}
Gilmore, R. (2006). Lie groups, Lie algebras, and some of their applications. Courier Corporation.

\bibitem{stone}
Stone, M. H. (1932). On one-parameter unitary groups in Hilbert space. Annals of Mathematics, 33(3), 643-648.

\bibitem{Magnus}
Magnus, W. (1954). On the exponential solution of differential equations for a linear operator. Communications on pure and applied mathematics, 7(4), 649-673.

\bibitem{Blanes}
Blanes, S., Casas, F., Oteo, J. A., \& Ros, J. (2009). The Magnus expansion and some of its applications. Physics reports, 470(5-6), 151-238.

\bibitem{pedag}
Blanes, S., Casas, F., Oteo, J. A., \& Ros, J. (2010). A pedagogical approach to the Magnus expansion. European Journal of Physics, 31(4), 907.

\bibitem{fard}
Ebrahimi-Fard, K., Mencattini, I., \& Quesney, A. (2023). What is the Magnus expansion?, Journal of Computational Dynamics, 2025, 12(1): 115-159.

\bibitem{iserl}
Iserles, A., \& Nørsett, S. P. (1999). On the solution of linear differential equations in Lie groups, Philosophical Transactions of the Royal Society of London, Series A: Mathematical, 357(1754), 983-1019.

\bibitem{casas}
Casas, F. (2007). Sufficient conditions for the convergence of the Magnus expansion. Journal of Physics A: Mathematical and Theoretical, 40(50), 15001.

\bibitem{mulian}
Mulian, Y. (2024). On the exact solution for the Schr\" odinger equation. Particles. 2024; 7(4):1095-1119.

\bibitem{nonhe1}
Bender, C. M. (2007). Making sense of non-Hermitian Hamiltonians. Reports on Progress in Physics, 70(6), 947.

\bibitem{nonhe2}
Moiseyev, N. (2011). Non-Hermitian quantum mechanics. Cambridge University Press.

\bibitem{nonhe3}
Bountis, T., \& Skokos, H. (2012). Complex hamiltonian dynamics (Vol. 10). Springer Science \& Business Media.

\bibitem{nonhe4}
Bagarello, F., \& Trapani, C. (2016). Non-Hermitian Hamiltonians in quantum physics (Vol. 184). New York: Springer.

\bibitem{nhse1}
Hatano, N., \& Nelson, D. R. (1996). Localization transitions in non-Hermitian quantum mechanics. Physical review letters, 77(3), 570.

\bibitem{nhse2}
Okuma, N., Kawabata, K., Shiozaki, K., \& Sato, M. (2020). Topological origin of non-Hermitian skin effects. Physical review letters, 124(8), 086801.

\end{thebibliography}
\end{document}